# Strain-tunable synthetic gauge fields in topological photonic graphene

*Zhen-Ting Huang[1], Kuo-Bin Hong[1], Ray-Kuang Lee[2,3], Laura Pilozzi[4,5], Claudio Conti[5,6], Jhih-Sheng Wu[\*,1] and Tien-Chang Lu[\*,1]*

[1] Department of Photonics and Institute of Electro-Optical Engineering, College of Electrical and Computer Engineering, National Yang Ming Chiao Tung University, Hsinchu 30050, Taiwan

[2] Institute of Photonics Technologies, National Tsing Hua University, Hsinchu 30013, Taiwan

[3] Physics Division, National Center for Theoretical Sciences, Hsinchu 30013, Taiwan

[4] Institute for Complex Systems, National Research Council (ISC-CNR), Via dei Taurini 19, 00185 Rome, Italy

[5] Research Center Enrico Fermi, Via Panisperna 89a, 00184 Rome, Italy

[6] Department of Physics, University Sapienza of Rome, Piazzale Aldo Moro 5, Rome 00185, Italy

*Correspondence and requests for materials should be addressed to:

timtclu@nycu.edu.tw (T.C.L.) and jwu@nycu.edu.tw (J.S.W.)

## Abstract

We propose a straightforward and effective approach to design, by strain-engineering, photonic topological insulators supporting high quality factors edge states. Chiral strain-engineering creates opposite synthetic gauge fields in two domains resulting in Landau levels with the same energy spacing but different topological numbers. The boundary of the two topological domains hosts robust time-reversal and spin-momentum-locked edge states, exhibiting high quality factors due to continuous strain modulation. By shaping the synthetic gauge field, we obtain a remarkable field confinement and tunability, with the strain strongly affecting the degree of localization of the edge states. Notably the two-domain design stabilizes the strain-induced topological edge state. The large potential bandwidth of the strain-engineering and the opportunity to induce the mechanical stress at the fabrication stage enables large scalability for many potential applications in photonics, such as tunable microcavities, new lasers and information processing devices, including the quantum regime.

# Introduction:

Topologically protected edge states with the immunity to distortions or fabrication imperfections of crystals have attracted much attention in recent years. Since the first discovery of the quantum Hall effect was reported [1], the conductive electronic state located at the structural boundary has become a popular topic in many electronic applications [2-5]. However, to induce the quantum Hall effect, a strong magnetic field is necessary. The field induces cyclotron motion and forms quantized electronic states called Landau levels in the bulk region. As a result, an edge state with broken time-reversal symmetry is presented in the system. In 2005, a breakthrough was discovered and proposed by Kane and Mele [6], who outlined that spin-dependent topological states without involving any external magnetic field may occur between K and K' point of the Brillouin zone in graphene. This notable feature pushes the topological edge state into practical applications as the pivotal building block for the development of topological insulators. Based on the superior transport properties of the massless Dirac fermions and topological protection, topological insulators have exhibited outstanding performance in topological superconductors [2, 7, 8], spintronic devices [3], and quantum computing [5]. In comparison to electrons, the coupling between the magnetic field and photons is weaker. Therefore, the spin-orbital coupling is usually used to create band inversion and change the photonic system's topology [9-11]. The corresponding topological edge state possesses the time-reversal symmetry and is protected from backscattering by spin-momentum locking [9, 12, 13], with inhibition of experimental fluctuations. Thanks to the unprecedented features, the photonic topological insulator has exhibited extraordinary performance in many optical devices, including unidirectional waveguides [14, 15], optical switching [16, 17], optical isolators [18], and lasers [19-21].

Mainly, most designs of topological photonics either utilized magnetic materials or phase resonators, which usually operate at low frequencies and lack of scalability. Interestingly, based on spin-orbital coupling, a synthetic gauge theory was proposed to mimic the quantum Hall effects in the absence of an external magnetic field [22]. The synthetic gauge field is created by the appearance of a continuously distributed strain, whose curl gives an artificial magnetic field. Notably, in the presence of the strain-induced pseudo-magnetic field, a photonic bandgap can be opened, and Landau levels appear in the band diagram of the photonic system [23-25]. These photonic Landau levels have a large group index and high density of state (DOS), leading to a strong Purcell effect and increasing the light-matter interaction. Additionally, the frequency gap between two Landau levels provides the insulating capability in the bulk region. According to the bulk-edge correspondence in the quantum Hall effect, there should be an edge state inside this frequency gap. However, only a few studies have successfully demonstrated the topological edge state under the pseudo-magnetic field [26]. Compared to other edge states constructed by breaking the parity symmetry [21, 27], the strain-induced edge state with an off-$\Gamma$ momentum shows a shorter propagation length due to significant diffraction loss near the structural boundary, which drastically limits its applicability [23, 26]. To make the strain-induced topological edge state stable enough to realize in the practical application, it is mandatory to identify a design that reduces diffraction loss.

Here, we propose a chiral structure obtained by a continuous displacement function in the arrangement of holes in a honeycomb superlattice. The displacement function corresponds to deformations sketched in Fig. 1. Fig. 1(a) shows a membrane with air holes arranged in a honeycomb lattice fixed at one end like a cantilever beam. An external force applied at the free-end produces a continuous deformation as shown in Fig. 1(b). Here we assume that the deformation induces a displacement but does not affect the shape of air holes. In Fig. 1(c), the strain of air holes in the membrane produces a synthetic vector potential, whose curl determines the strength and direction of the pseudo-magnetic field. Remarkably, when increasing the strength, a local bandgap is opened and gradually broadened at the first Dirac cone of the honeycomb lattice (Fig. 1(e)), and Landau levels with different valley Chern numbers ($C$) form at each K and K' valley in the band diagram (Fig. 1(f)) [28], as in the quantum Hall effect but still preserving the time-reversal symmetry. Accordingly, the strain-induced pseudo-magnetic field must be reversed between K and K' valley to maintain the time-reversal symmetry in the global Dirac Hamiltonian [29]. Fig. 1(d) shows a chiral structure composed of two strained patterns with the pseudo-magnetic field in opposite directions. Owing to the reverse topology resulting from the opposite pseudo-magnetic fields, topological edge states appear between different orders of Landau levels (Fig. 1(g)), where the red and green solid line represents the spin-up and spin-down edge state, respectively. These modes are spatially located at the boundary between the two domains. Furthermore, if the direction of lattice arrangement perpendicular to the external force shown in Fig. 1(b) is along the zigzag, the bands at K and K' valley after the deformation would be simultaneously projected to the $\Gamma$ point based on Bloch's band theory. Subsequently, the crossing bands, which include the spin-up edge state in one valley and spin-down edge state in the other valley, appear between Landau levels and belong to the radiation mode because these bands are in the light cone, as shown in Fig. 1(h). It's worth noting that tuning the magnitude and peak position of the pseudo-magnetic field, a highly localized strain-induced topological edge state at $\Gamma$ point can be obtained.

The mode localization of strain-induced topological edge states is significantly enhanced because of the continuous modulation, which implies the diffraction loss of the strain-induced topological edge states can be extremely reduced with respect to the edge state of the 0$^{th}$ Landau level [26]. This method provides an effective way to design a lossless edge state, which also exhibits scattering-free and unidirectional characteristics. Additionally, strain-engineering applies to all the frequencies and is suitable for photonics since strains can be arranged on purpose at the fabrication stage. With such excellent performance, our results pave the way for the practical application of strain-induced topological edge states and are beneficial for low-energy-consumption optical devices.

**Theories and Methods:**

The unstrained system is made of a GaAs membrane with air holes disposed in a honeycomb lattice with an armchair boundary. As shown in Fig. 2(a), the strained system is obtained by adding a continuous displacement function to the arrangement of air holes. Given the strain ($\varepsilon$)-displacement (u) relation:

$$\begin{pmatrix} \varepsilon_{xx} & \varepsilon_{xy} \\ \varepsilon_{yx} & \varepsilon_{yy} \end{pmatrix} = \frac{1}{2} \begin{pmatrix} 2u_{x,x} & u_{x,y} + u_{y,x} \\ u_{y,x} + u_{x,y} & 2u_{y,y} \end{pmatrix} \quad (1)$$

the resulting strain distribution for a given displacement function can be obtained. Then, according to the work published by F. Guinea in 2009 [30], this strain can generate a gauge field A and lead to the Landau quantization at K valley [31-35],

$$\begin{pmatrix} A_x \\ A_y \end{pmatrix} = \frac{\beta}{a} \begin{pmatrix} \varepsilon_{xx} - \varepsilon_{yy} \\ -2\varepsilon_{xy} \end{pmatrix} \quad (2)$$

where $\beta = -\partial \ln t / \partial \ln a$, $t$ is the hopping parameter, and $a$ is the lattice constant. The $\sqrt{3}a$ is chosen as 350 nm in our honeycomb lattice. Hereafter, the pseudo-magnetic field can be determined by calculating the curl of the synthetic gauge field. To analogize the quantum Hall effect, an out-of-plane magnetic field is necessary. Therefore, we consider only its z-component, which implies that $A_{x,y}$ and $A_{y,x}$ will dominate the strain-induced quantum Hall effect. Besides, the edge state in Fig. 2(a) is designed to propagate along the *y*-direction, with the lattice preserving the periodicity in the same direction. Thus, we only add an *x*-varied displacement into our system, which means the $A_{x,y}$ term is null. Finally, based on equation (1) and (2), the relation between the displacement and pseudo-magnetic field can be derived as follows:

$$B_z = A_{y,x} = -\frac{\beta}{a} u_{y,xx} \qquad (3)$$

In equation (3), the distribution of the pseudo-magnetic field only depends on the second-order differentiation of $u_y$. If an inflection point is added to the input displacement function, the structure will be divided into two regions. The region with a positive second derivative of the displacement exhibits the pseudo-magnetic field in the +*z*-direction, and the other region exhibit the pseudo-magnetic field in the –*z*-direction. The strain-induced quantum Hall effect in these two regions will display opposite characteristic and then result in a reverse topology. As a consequence, the boundary defined by the inflection point host topological edge states, which are protected from the diffraction loss near the structural boundary, giving, a chiral structure with tunable strong localization as designed in Fig 2(a) following the approach by F. Guinea [30]. However, the strain patterns in [30] seems unlikely to generate propagating edge states. On the contrary, our approach allows high Q edge modes with also tunable localization widths. All the simulations were calculated by using the frequency domain solver of the finite-element software (COMSOL Multiphysics)

## Results:

To induce a pseudo-magnetic field, a continuous displacement as a function of the *x*-coordinate in the *y*-direction is applied to the suspended GaAs membrane. The schematic diagram of the displacement function is shown in Fig. 2(b). First, we focus on a Lorentzian-distributed pseudo-magnetic field:

$$B_z(x) = \frac{\beta}{a} \frac{b^2}{(x-L)^2 + b^2} \tag{4}$$

where $L$ and $b$ signify the peak position and the half-width at half-maximum (HWHM), respectively. The Lorentzian-distributed pseudo-magnetic field would give the local response of the strain-induced quantum Hall effect. Then, according to equation (3), the displacement function can be inversely derived by substituting some boundary conditions, including fixing the maximum displacement $u_{max}$ at $x=0$, no displacement at $x=R$, and the first derivative of the displacement equal 0 at $x=R$, where $R$ represents the final $x$-coordinate in Fig. 2(b). Afterward, a continuous displacement function with a Lorentzian-distributed pseudo-magnetic field can be obtained. Figure 2(c) demonstrates the derived displacement function and its second derivative on the condition that fix $u_{max}$ and $L$ as 500 nm and 0.5×$R$, respectively.

I. **Strain-induced Landau levels**

Based on the derived distribution of the displacement, the band diagram varied with $u_{max}$ is shown in Fig. 3. The honeycomb supercell in the simulation model is fixed as 60 periods in the $x$-direction, as shown in Fig. 2(b). When $u_{max}$ is increased in the displacement function, the peak of the second derivative is raised, resulting in an enhancement of the pseudo-magnetic field. Therefore, the strain-induced quantum Hall effect should become more significant as enlarging $u_{max}$. Figure 3(a) shows the band diagram in the absence of the input displacement. Although there is no shift in the arrangement of air holes, a small bandgap still appears at the Γ point, which is due to the finite number of periods in the $x$-direction. As the displacement is added, and $u_{max}$ is increased to 200 nm, the local bandgap slightly broadens, and three photonic Landau levels indicated as red dot lines with a higher DOS appear in the band diagram, as shown in Fig. 3(b). When $u_{max}$ continues to be increased to 500 nm, the local bandgap at the Γ point is further expanded, which is as large as 15 nm observed in Fig. 3(c). Most vitally, increasing $u_{max}$ up to 500 nm in the displacement function implies the Lorentzian-distributed pseudo-magnetic field becomes sharper in the spatial distribution, causing a significant local response of the strain-induced quantum Hall effect and an increase in the energy splitting $\Delta E_g$ between photonic Landau levels, as shown in Fig. 3(d). This striking phenomenon can be also observed in the variation of DOS of photonic bands in the band diagram. Figure 3(e) shows the electric field distribution of the three photonic Landau levels at the Γ point, and the corresponding valley Chern number is also indicated in each figure. The detailed discussion of the valley Chern number is demonstrated in the first section of the Supplementary Information. Notably, the mode profile of each photonic Landau level is similar at the Γ point and mainly distributed on the right side of the structure, which is close to the undeformed structural boundary shown in Fig. 2(c). According to the nature of Landau levels, the wave function will be changed from the bulk state to the edge state as the $k$ along the periodic direction gradually deviates from the reciprocal lattice point [25], which is the Γ point in this work. However, because of the asymmetric deformation we applied, the bulk state at the Γ point also shows the broken geometric symmetry in each photonic Landau level, whose electric field is blocked at the peak position of the pseudo-magnetic field. To verify more detailed features, the electric field distribution of the $0^{th}$ photonic Landau level at several off-Γ points is further extracted and shown in Fig. 3(f). It is obvious to see that when $k_y$ is increased and deviated from the Γ point, the electric field gradually shifts to the left side of the structure, where the strain is primarily distributed. This striking phenomenon demonstrates the state conversion of the photonic Landau level and becomes the apparent evidence to prove the appearance of strain-induced Landau levels. According to these simulation results, even if the pseudo-magnetic field is non-uniform, photonic Landau levels still exist in the band diagram. The appearance of photonic Landau levels and the size of the local bandgap is strongly related to the magnitude of the pseudo-magnetic field, which satisfies the phenomenon of the quantum Hall effect. On top of that, the Lorentzian-distributed function gives a large flexibility to tune the peak position and maximum strength of the pseudo-magnetic field and exhibits tunable capability in photonic gaps by strain-engineering.

## II. Topological edge state in the chiral structure

After demonstrating photonic Landau levels, we can go further to realize the topological edge state. An idea to produce the topological edge state with an ultra-strong mode localization is to introduce an inflection point into the displacement function. Therefore, the displacement in Fig. 2(c) is duplicated and then rotated 180 degrees related to the coordinate: ($x=R$, $y=\sqrt{3}a/2$) to constitute the displacement distribution with an inflection point, whose schematic diagram of the combination and the corresponding second derivative are shown in Fig. 4(a) and (b), respectively. In the light of the derivation in equation (3), the two combining structures separately produce the positive and negative pseudo-magnetic field, which leads to the reverse strain-induced quantum Hall effect and opposite topologies in photonic bands. Consequently, through this structural combination, a chiral pattern is constructed, and the combining boundary located at $x=R$ is regarded as the demarcation line of two reverse quantum Hall effects. Figure 4(c) shows the band diagram of the chiral structure. It is astonishing to see that two crossing bands appear between the $0^{th}$ and $C=\pm1$ photonic Landau levels. Base on the discussion in Fig. 1, it can be intuitive to understand that these crossing bands are constituted by topological edge states at K and K' valley. Additionally, photonic Landau levels near the Γ point disappear in the band diagram, which is because the right boundary in Fig. 2(b) is vanishing after the structural combination. As $k_y$ is gradually shifted from the Γ point, photonic Landau levels emerge again owing to its state conversion, and the corresponding electric fields are distributed in the large strain region, as shown in the lowest figure of Fig. 4(d). For the lower crossing band, the corresponding electric field distributions of topological edge state σ+ and σ- indicated at point σ+ and σ- in Fig. 4(c) are shown in Fig. 4(d). Notably, these states are localized at the combining boundary, which is beneficial for reducing the diffraction loss. Most crucially, to further discuss the topological characteristics, the correlated in-plane electric field vector and Poynting vector are also demonstrated in the distribution of black arrows in Fig. 4(e) and (f), respectively. The governing equation we solved in the simulation model is simplified in the transverse electric (TE) form, so the dominated component of the magnetic field $H$ is in the z-direction, and the $H_z$ distributions of topological edge state σ+ and σ- in the black-dash box both marked in Fig. 4(d) are also shown in Fig. 4(e) and (f). Apparently, by observing the corresponding polarization and power flow, the photonic spin-up state can be analogized to topological edge state σ+ owing to the left-hand circular polarization (LCP), and its propagation direction is in the +y-direction. On the contrary, topological edge state σ- is the right-hand circular polarization (RCP), which can also analogize the photonic spin-down state, and the propagation direction is in the –y-direction. These results have exhibited robust evidence for the strain-induced topological edge state in spin-momentum locking. Furthermore, the angular momentum $J$ has been introduced to calculate the effect of the strain-induced pseudo-magnetic field, whose equation is demonstrated in the following [36]:

$$\vec{J} = \int \frac{\vec{r} \times (\vec{E} \times \vec{B})}{4\pi c} dA \qquad (5)$$

where *r* and *c* represent the spatial coordinate and speed of light, respectively. By introducing equation (5), the angular momentum of topological edge state σ+ and σ- are separately calculated as a positive and negative value in the *z*-direction, whose sign is determined by the LCP or RCP of the topological edge state. Most vitally, pivotal evidence to prove that the edge state can be only created by the specific structural combination and the detailed calculated value of *J* are also discussed in the Supplementary Information. Figure s1(a) shows the displacement distribution after the structural combination shown in Fig. 4(a) and (b), which is composed of the positive and negative pseudo-magnetic field but setting *L* as R in equation (3). Besides, a displacement distribution composed of two positive pseudo-magnetic fields is constructed and shown in Fig. s1(b). The corresponding band diagrams of the structures arranged by the displacement function in Fig. s1(a) and (b) are shown in Fig. s2(a) and (b), respectively. Obviously, only the constitution of the positive and negative pseudo-magnetic field can support the topological edge state, which proves that the strain-engineering we propose is an effective way to produce the opposite pseudo-magnetic fields and then reverse the band topologies.

### III. The tunable ability of the strain-induced pseudo-magnetic field

According to the quantum Hall effect in electronics, the energy splitting between Landau levels is proportional to the external magnetic field. When a strong magnetic field is applied, the large energy gap provides an extraordinary insulating capability in the bulk region, causing the topological edge state to be highly localized at the structural boundary. As to the strain-engineering synthetic gauge field, if $u_{max}$ in Fig. 2(c) is increased, the local $u_{y,xx}$ is increased too, resulting in a significant enhancement of the pseudo-magnetic field based on equation (3). Therefore, the strength of the pseudo-magnetic field is mainly dependent on $u_{max}$. To verify the strain-induced quantum Hall effect, $u_{max}$ in the chiral structure shown in Fig. 5(a) is tuned to observe the variation of the localization length in the topological edge state. The equation of the localization length derived from the Anderson localization is shown in the following [37-39]:

$$\text{Localization length} = [\frac{\int |E_{avg,lowpass}|^4 \, dx}{(\int |E_{avg,lowpass}|^2 \, dx)^2}]^{-1} \qquad (6)$$

where $E_{avg,lowpass}$ is the electric field averaged along the y-direction in a supercell after passing a low-pass filter, and the detailed calculation is discussed in Fig. s4. In addition, the parity is discussed to explore the relation between the two topological edge states. Figure 5(b) shows $H_z$ of topological edge state σ+ and σ- as the function of the x-coordinate, which is also calculated by averaging along the y-direction in a supercell, and the enlarged blue dash box is also shown in Fig. 5(c). It is clear to see that topological edge state σ+ and σ- have the opposite parity, which implies these states are correlated with each other. Afterward, the $u_{max}$-dependent $|E_{avg}|^2$ distribution of topological edge state σ+ is demonstrated to discover the strength effect of the pseudo-magnetic field in Fig. 5(d). As $u_{max}$ is increased, the topological edge state is gradually confined to the combining boundary (x=R). To further quantify the increased confinement, the localization length is also calculated by equation (6) and shown in Fig. 5(e). Remarkably, the increment of $u_{max}$ can dramatically enhance the mode localization, and the localization length is varied from 17.5 μm to 9.69 μm as increasing $u_{max}$ from 100 nm to 1000 nm. The strong localized feature makes the continuous strain modulation exhibit the great potential to reduce the diffraction loss near the structural boundary and improve the stability of topological edge states.

Furthermore, the mode localization is sensitive to the peak position of the pseudo-magnetic field. Figure 6(a), (b), and (c) shows the displacement distribution as setting L as 0, R/2, and R in equation (4). If the peak position of the local responded pseudo-magnetic field is gradually far from the combining boundary, the mode profile of the topological edge state will be extended owing to the awful insulating capability in the bulk region. Consequently, as $L$ is closed to the combining boundary as shown in the $L$-dependent $|E_{\text{avg}}|^2$ distribution in Fig. 6(d), the mode is rapidly converged to the center position, and the localization length in Fig. 6(e) is shrunk as low as 7.45 μm when $L$ is up to $R$. More interestingly, $\Delta E_g$ is also affected by the peak position of the pseudo-magnetic field, as shown in the red line of Fig. 6(e), which shows that the strain-induced quantum Hall effect becomes more significant as the pseudo-magnetic field approaches the boundary. The design method proposed here provides an unprecedented way to architect a tunable topological edge state with the capability to achieve an ultra-strong localization, which also exhibits the great potential to be realized in practical application.

## Conclusion:

We studied topological photonic structures constituted by a chirally strained honeycomb lattice. We show that strain induces tunable gauge fields, which create Landau levels. At the boundary of two domains with opposite pseudo magnetic fields, robust topological edge states appear at the boundary. The edge states correspond to crossings in the original band gaps. On the contrary, no edge states appear at the boundary of two domains with pseudo magnetic fields of the same sign. The edge states have topological origin and are spin-momentum locked. At variance with other strain-induced topological photonic designs, smooth chiral strain-engineering gives rise to strongly localized edge states with tunable capability. We showed that the confinement of the edge states is enhanced by modifying local strain distribution. We also quantify the corresponding increment of the degree of mode localization. Surprisingly, the localization length of the edge states can be reduced below 7.45 μm when choosing the $L/R$ as 1 and $u_{\text{max}}$ above 500 nm in equation (6). Hence, strongly localized topological edge states can be observed when the peak position of an intense pseudo-magnetic field is at the boundary of two domains. In photonic systems, strain can be induced during fabrication, therefore, controlling the lattice displacement affects the strength of modulation and produces a pseudo magnetic field. This approach applies to photonic crystals regardless of their frequencies and sizes and hence enables broadband applications.

The proposed chiral strain-engineering topological photonics has excellent potential in advanced sensing and other applications. One can imagine innovative information processing devices, as filters and resonators, for passive and active tunable devices, including lasers and quantum optical systems.

## Acknowledgements

This work has been supported in part by the Higher Education Sprout Project of the National Yang Ming Chiao Tung University and Ministry of Education (MOE), Taiwan, and in part by the Ministry of Science and Technology in TAIWAN under Contract No. MOST 110-2218-E-A49-012-MBK, MOST 109-2627-M-008-001 and MOST 110-2221-E-A49 -058 -MY3.

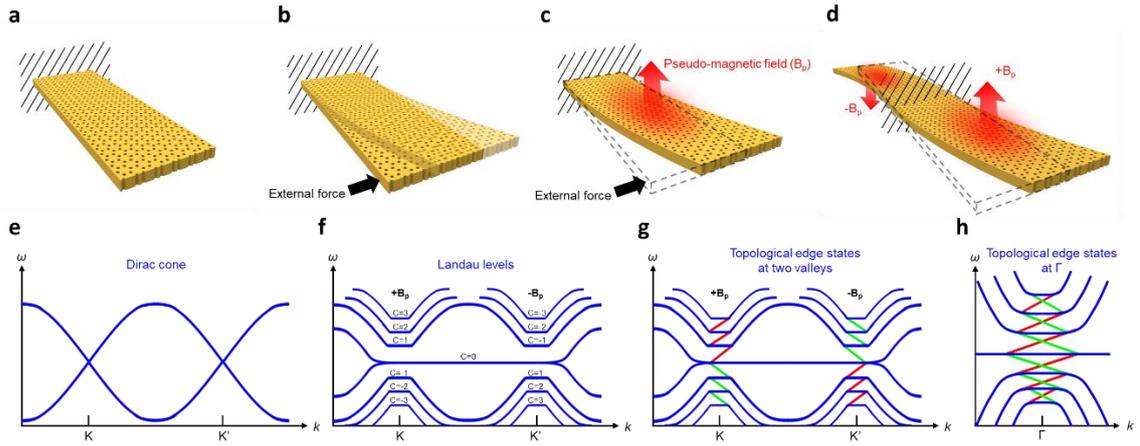

**Fig. 1** | The schematic diagrams of deformation processes of a membrane containing honeycomb superlattice. (a) Before applying an external force, a membrane with a honeycomb lattice is fixed like a cantilever beam. (b) After applying an external force, the membrane is deformed along the applying direction of the force with one end fixed. (c) A synthetic vector potential is produced by the continuously distributed strain, and its curl can define a pseudo-magnetic field $B_p$. (d) A chiral structure, which is composed of two patterns with opposite pseudo-magnetic fields. (e) The band diagram of the membrane before the deformation, and two Dirac cones exist at the K and K' valley. (f) After the deformation, bands split into several Landau levels, and a local bandgap appears in the band diagram. Also, the corresponding valley Chern numbers ($C$) are indicated in this figure, and the central flat band represents the $0^{th}$ photonic Landau level. (g) The band diagram of two opposite chiral structures. After the combination of two patterns with opposite pseudo-magnetic fields, topological edge states appear between each Landau level at the K and K' valley, where the red and green solid lines are indicated as the spin-up and spin-down edge states respectively. (h) Topological edge states at the K and K' valley are simultaneously projected to the Γ point based on Bloch's band theory because the direction of lattice arrangement along the long axis of the membrane is along the zigzag.

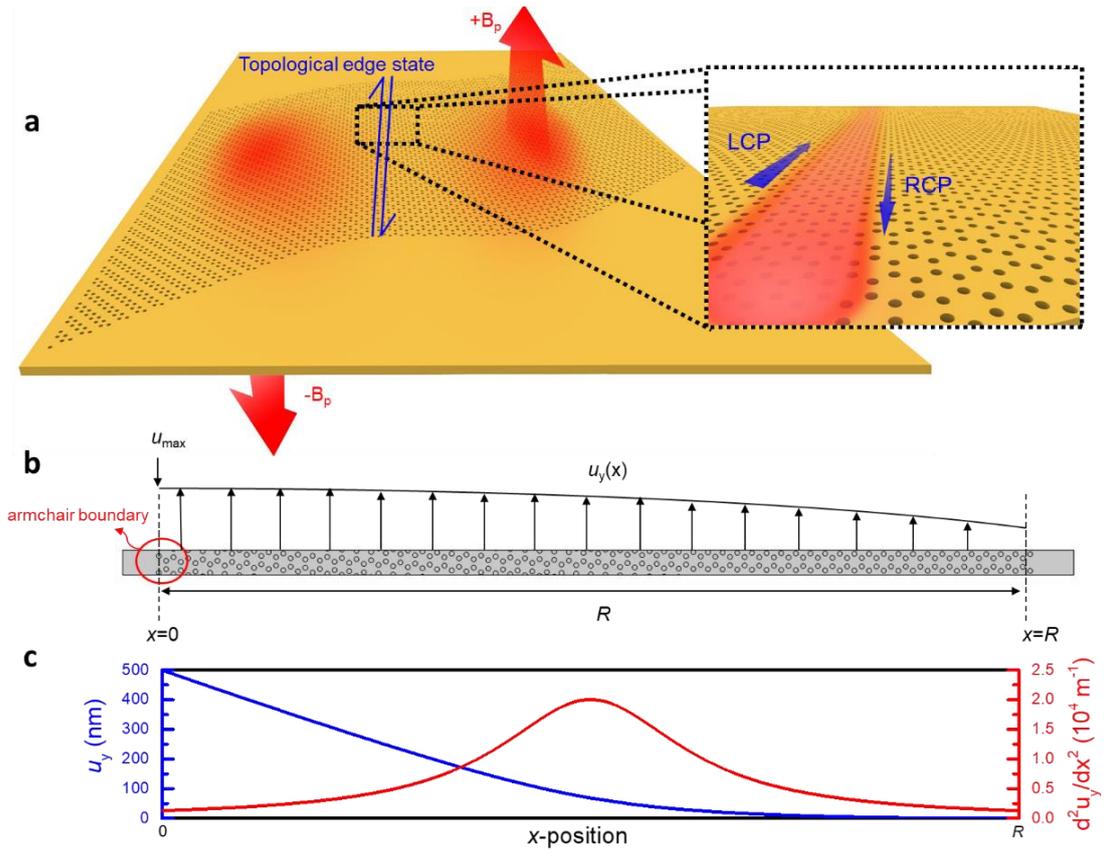

**Fig. 2** | The schematic diagrams of the proposed chiral structure and the applying displacement. (a) The schematic diagram of the chiral structure, which supports the strain-induced topological edge state. (b) The schematic diagram of the honeycomb superlattice and the continuous $y$-displacement, which is the function of $x$ and distributes from $x=0$ to $x=R$, and the $x$-axis is along the zigzag direction of the honeycomb lattice. (c) The continuous displacement functions. The blue line indicates the distribution of $u_y$, and the red line indicates the distribution of $\partial^2 u_y / \partial x^2$.

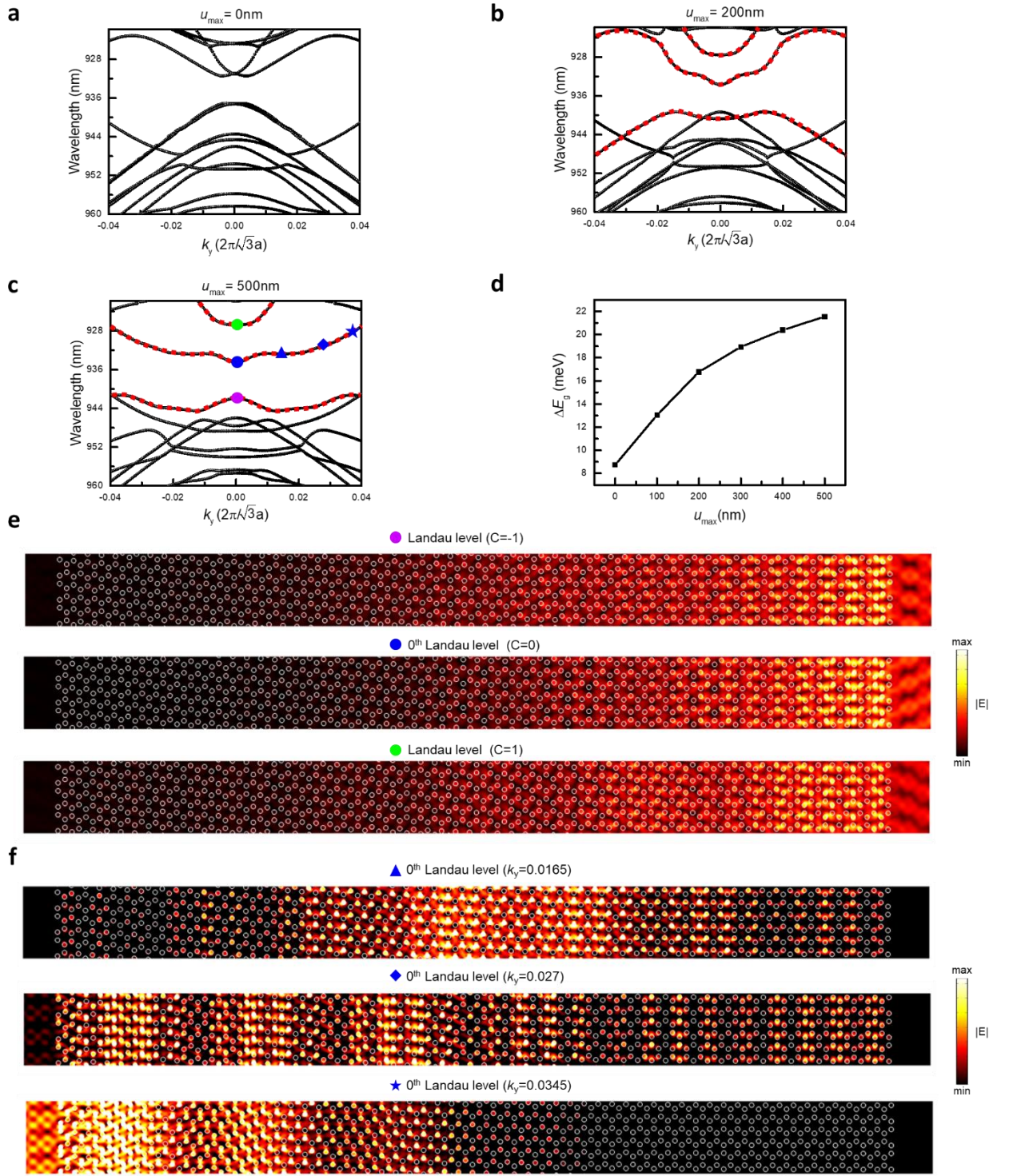

**Fig. 3 |** The appearance of photonic Landau levels. (a)-(c) The band diagram when $u_{max}$ equals 0 nm (a), 200 nm (b), and 500 nm (c) in equation (1), respectively. The red dot lines in b and c indicate the photonic Landau levels. (d) $\Delta E_g$ as the function of $u_{max}$. (e) The electric field distribution of the three photonic Landau levels at the Γ point and the corresponding valley Chern number. (f) The electric field distributions of the 0$^{th}$ photonic Landau level at 0.0165, 0.027, and 0.0345 of $k_y$ in c.

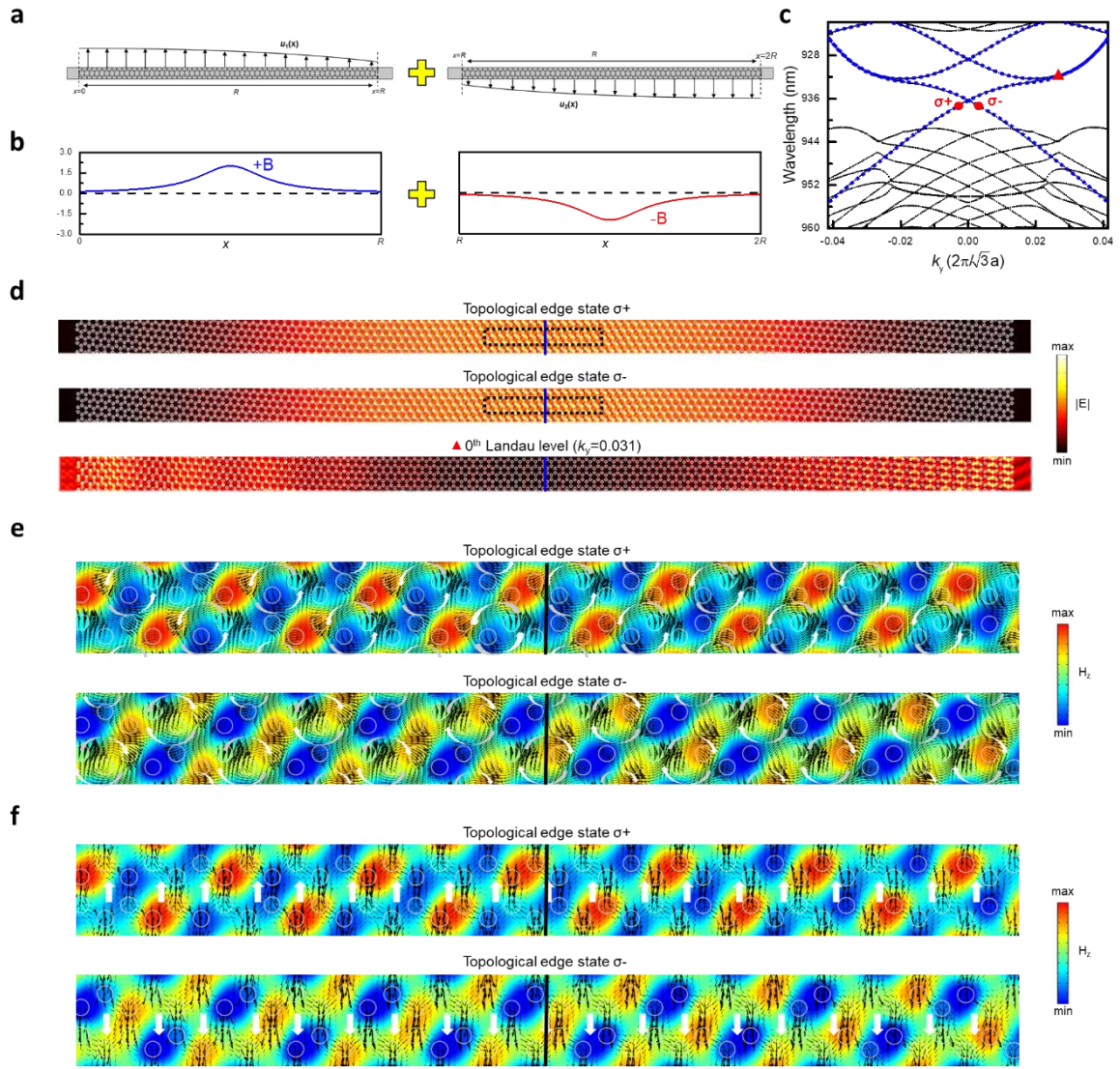

**Fig. 4** | Strain-induced topological edge states. (a) The schematic diagram of the combination of the chiral structure, which is composed of two strained patterns with (b) the pseudo-magnetic field in opposite directions. (c) The band diagram of the chiral structure, which fixes $u_{max}$ as 500 nm in equation (1). (d) The corresponding electric field distribution of topological edge state σ+ and σ- signed as circular points in c, and the electric field distribution of the 0$^{th}$ photonic Landau level at 0.031 of $k_y$ signed as a triangular point in c. (e) The distribution of $H_z$ component, where the black arrows indicate the distribution of the in-plane electric field. (f) The distribution of $H_z$ component, where the black arrows indicate the distribution of the Poynting vector.

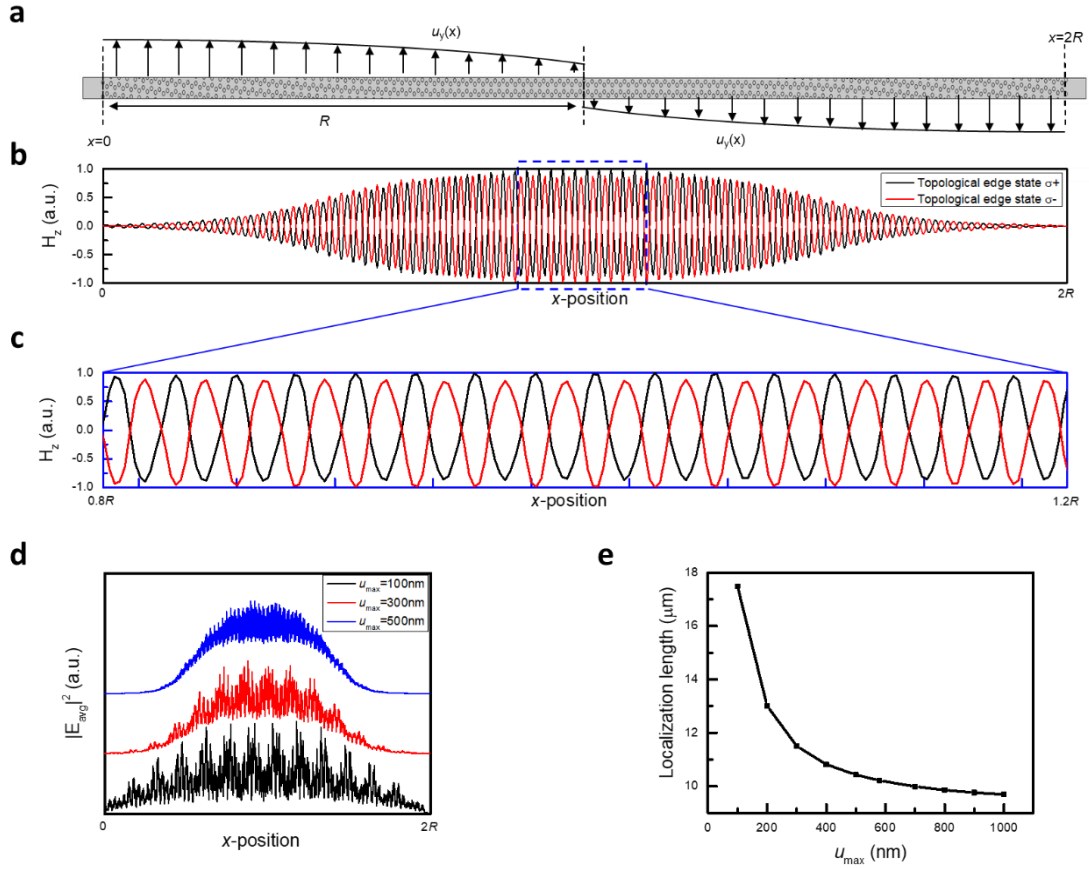

**Fig. 5** | The parity of topological edge states and the tunability based on the strength of the pseudo-magnetic field. (a) The supercell of the chiral structure, which has Ander periods in the *x*-direction and maintains periodic in the *y*-direction. (b),(c) The distribution of $H_z$ component in the *x*-direction, which is averaged along the *y*-direction in a supercell. The black and red line represent the edge state σ+ and σ- shown in Fig. 3, respectively, and the enlarged blue dash box is shown in c. (d) The $|E_{avg}|^2$ of edge state σ+ in $u_{max}$=100 nm, 300 nm, and 500 nm. (e) The mode localization as the function of $u_{max}$.

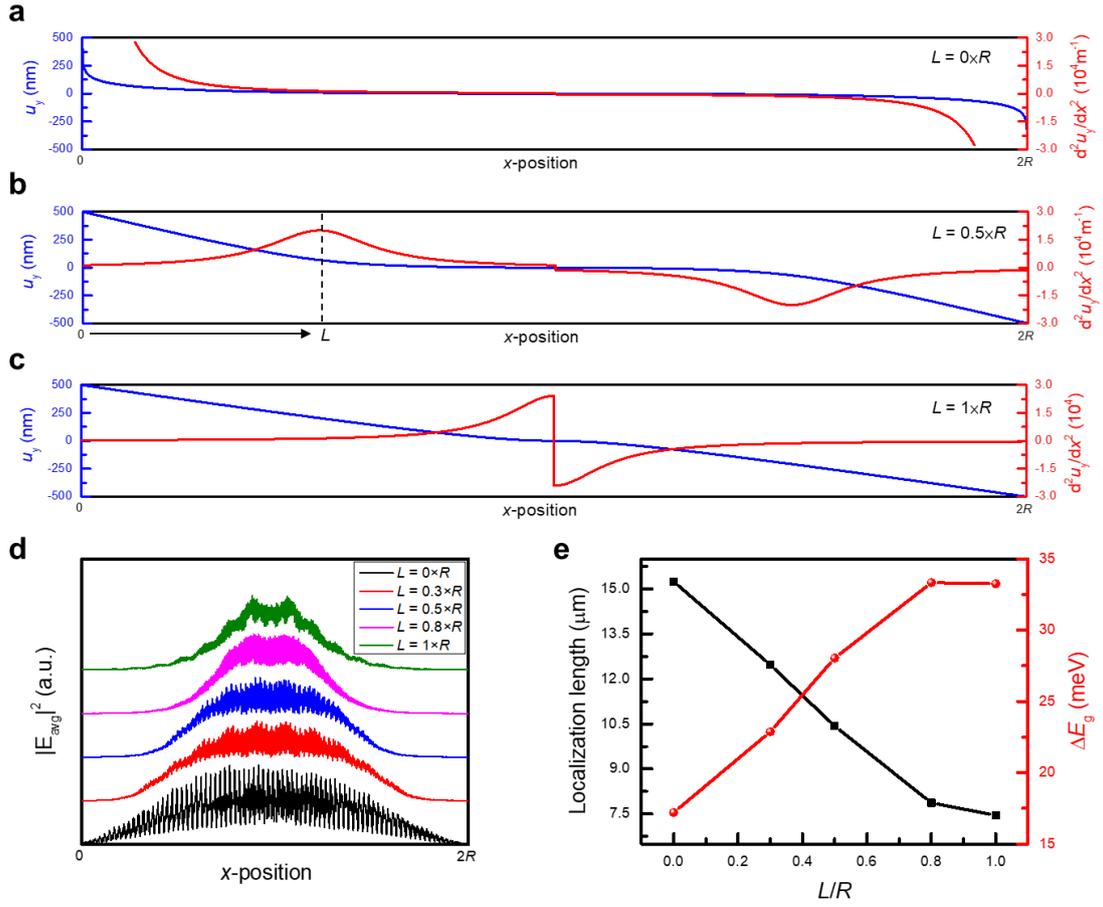

**Fig. 6 |** The tunability based on the peak position of the pseudo-magnetic field. (a)-(c) The continuous displacement function in different peak position of the pseudo-magnetic field, including $L=0$ (a), $0.5 \times R$ (b), and $1 \times R$ (c). (d) The $|E_{\text{avg}}|^2$ of topological edge state σ+ in $L=0 \times R$, $0.3 \times R$, $0.5 \times R$, $0.8 \times R$, and $1 \times R$. (e) The mode localization and $\Delta E_g$ as the function of $L/R$.

# Supplementary Information

## ■ The valley Chern number in each photonic Landau level

As the strain-induced pseudo-magnetic field is operated, the Hamiltonian in the strained system can be regarded as equation (s1), which is derived by substituting the Bloch condition to the Hamiltonian in Dirac Landau levels [1].

$$H = hv \begin{pmatrix} 0 & k_y - \partial_x + \dfrac{eB_p}{\hbar} x \\ k_y + \partial_x + \dfrac{eB_p}{\hbar} x & 0 \end{pmatrix} \qquad (s1)$$

$B_p$ is the strain-induced pseudo-magnetic field as discussed in the main text. Afterward, modified eigenvalues and eigenvectors can be solved, and then the Berry connection and Berry curvature can be further derived by equations (s2) and (s3) [2,3].

$$A_c(k) = -i <\psi(k)|\nabla_k|\psi(k)> \qquad (s2)$$

$$\Omega(k) = \nabla_k \times A_c(k) \qquad (s3)$$

Notably, when integrating the Berry curvature along the full Brillouin zone boundary, the Chern number can be determined but must equal zero owing to the time-reversal symmetry. However, in each valley, the integration of Berry curvature is nonzero, and the valley Chern number can be decided by counting the number of peaks of the Berry curvature [1]. In this work, finite-element software (COMSOL Multiphysics) is applied to solve the wave function in the strained structure. After calculating the Berry curvature near the K or K' valley and counting the peak number, the valley Chern number in each photonic Landau level is determined.

◼ **The verification of topological edge states in the chiral structure**

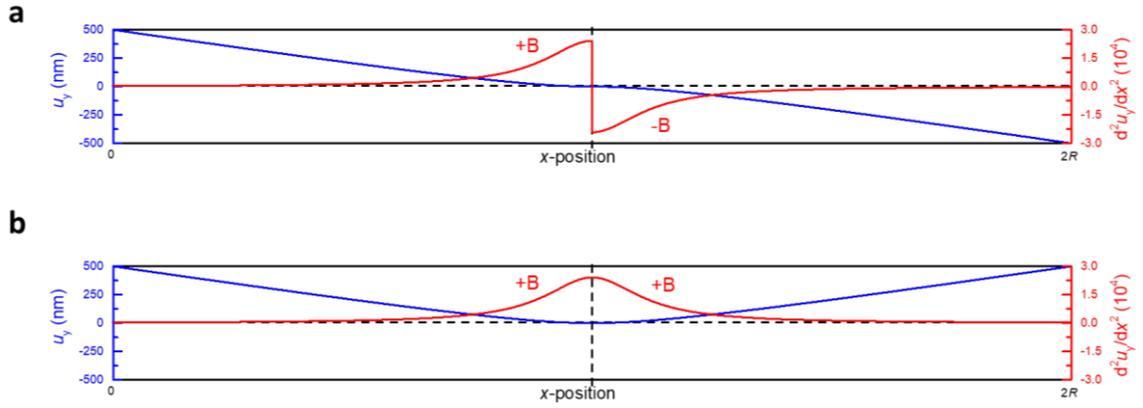

**Fig. s1** | The verification of topological edge states. (a),(b) The continuous displacement function after the structural combination of positive/negative (a) or positive/positive (b) pseudo-magnetic field, where $u_{max}$ and $L$ are set as 500 nm and $1\times R$, respectively. The blue line indicates the distribution of $u_y$, and the red line indicates the distribution of $\partial^2 u_y / \partial x^2$.

To verify the formation mechanism of topological edge states induced by our proposed chiral structure, two displacement functions are considered and shown in Fig. s1. First of all, Fig. s1(a) shows the displacement distribution after a combination of two patterns with positive and negative pseudo-magnetic field, which is the same as Fig. 6(c) in the main text. The opposite sign of the pseudo-magnetic field causes the reverse quantum Hall effect in the chiral structure, leading to a reverse topology in photonic Landau levels. Then, the topological edge state would exist between the two combining patterns. Figure s1(b) shows another kind of structural combination, which is composed of two patterns both with the positive pseudo-magnetic field. Owing to the same sign of pseudo-magnetic field, the topology of photonic Landau levels in two combining patterns is the same. Consequently, there should be no topological edge state located at the central boundary. By calculating band structures in these two structural combinations, the formation of our proposed strain-induced topological edge states can be verified. Furthermore, the photonic angular momentum introduced in equation 5 in the main text is also calculated to ensure the topological characteristics. If edge states in the chiral structure belong to the topological edge state, the spin-momentum locked condition must be satisfied. Namely, the angular momentum of edge states should be in the opposite direction, which is owing to the reverse spin and orbital propagation. In our calculation, the angular momentum of topological edge state σ+ and σ- in Fig. 4 are separately calculated as $5.282\times10^{-12}$ J·s and $-5.129\times10^{-12}$ J·s in the *z*-direction, which verifies the topological nature.

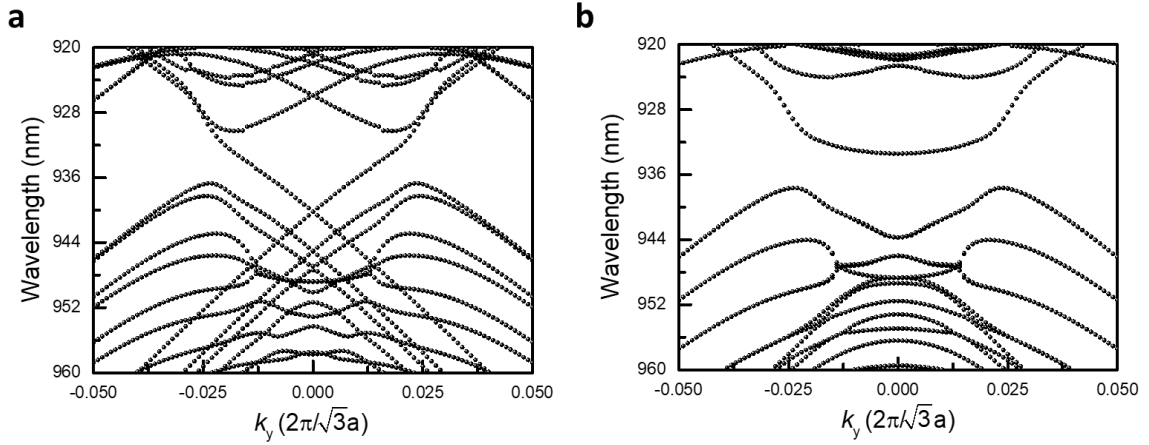

**Fig. s2** | The verification of topological edge states. (a),(b) The band diagram of the chiral structure, which is obtained by combining two patterns giving positive/negative (a) or positive/positive (b) pseudo-magnetic field.

Figure s2(a) and (b) demonstrate the corresponding band diagram of the structure with the displacement shown in Fig. s1(a) and (b), respectively. Notably, in Fig. s2(a), there are two crossing bands inside the local bandgap, located between photonic Landau levels, which satisfy the characteristic of strain-induced topological edge states discussed in the main text. On the contrary, in Fig. s2(b), there is only a flat band inside the local bandgap, which belongs to a defect mode produced by the combination of two patterns. Although its spatial distribution is also located near the combining boundary, this mode cannot propagate along the *y*-direction, a feature inferable by its slope shown in the band diagram. In this paper, we use a finite-element software (COMSOL Multiphysics) to perform the simulation. According to the finite-element method, the degrees of freedom number created by the cutting mesh will determine the number of eigenvalues observed in the simulation. Therefore, some virtual bands, which define as tricky solutions and cannot be observed in the experiment due to the ultra-low quality factor, appear in the calculated band diagram, as shown in the red and dense curves in Fig. s3(a). To simplify our calculation, we use some methods introduced in Fig. s3 to filtrate these unwanted bands.

# The filtration of virtual bands in the finite-element method

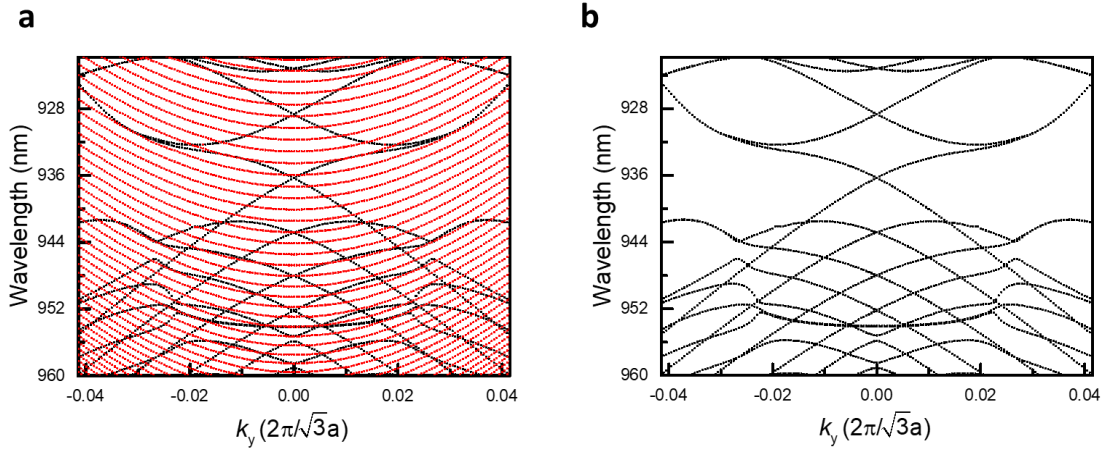

**Fig. s3** | The filtration of virtual bands. (a),(b) The calculated band diagram before (a) and after (b) the filtration of virtual bands. The red bands in a indicate the calculated virtual bands.

Those virtual bands appearing in the band structures can be separated from our simulation through a quality factor filter because these bands show ultra-low quality factors. Figure s3(a) is the calculated band diagram shown in Fig. 4(c) in the main text. Before filtrating, the solutions show many bands with ultra-low quality factors inside the local bandgap, making it difficult to observe the band-crossing feature of topological edge states. After the filtration, the two striking crossing modes clearly appear inside the local bandgap, as shown in Fig. s3(b). These virtual bands are tricky and hard to be measured in the experiments. Consequently, we remove these modes in the main text.

## ■ The calculation of localization length

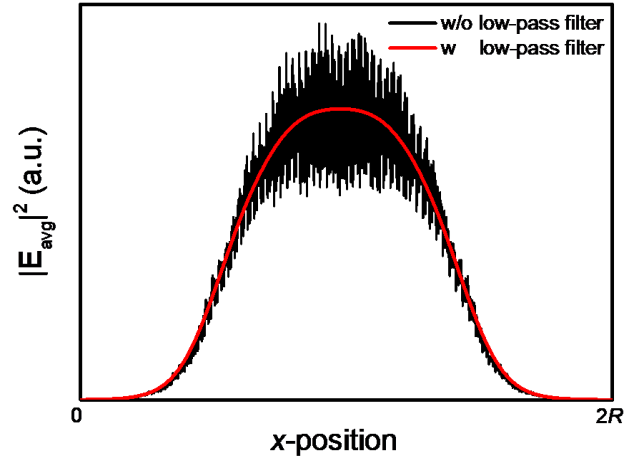

**Fig. s4** | The calculation with a low pass filter. The $|E_{avg}|^2$ distribution of topological edge state σ+ when $u_{max}$=500 nm and $L$=0.5×$R$.

To estimate the localization length introduced in equation 6 in the main text, a low-pass filter is used to obtain $E_{avg,lowpass}$ by suppressing the fluctuation resulted from the sampling accuracy. Figure s4 is the $|E_{avg}|^2$ distribution of topological edge state σ+ mentioned in Fig. 4(c) with and without a low-pass filter, which only allows passing the fundamental frequency. After the low-pass filter, we can obtain a clear distribution of $E_{avg,lowpass}$.